\documentclass[9pt,conference]{sig-alternate-05-2015}

\usepackage[font=scriptsize,skip = 1.5pt,labelfont=bf]{caption}
\usepackage{cite}
\begin{document}

% Copyright
\setcopyright{acmcopyright}
%\setcopyright{acmlicensed}
%\setcopyright{rightsretained}
%\setcopyright{usgov}
%\setcopyright{usgovmixed}
%\setcopyright{cagov}
%\setcopyright{cagovmixed}

% DOI
%\doi{10.475/123_4}

% ISBN
%\isbn{123-4567-24-567/08/06}

%Conference
%\conferenceinfo{PLDI '13}{June 16--19, 2013, Seattle, WA, USA}

%\acmPrice{\$15.00}

%
% --- Author Metadata here ---
%\conferenceinfo{WOODSTOCK}{'97 El Paso, Texas USA}
%\CopyrightYear{2007} % Allows default copyright year (20XX) to be over-ridden - IF NEED BE.
%\crdata{0-12345-67-8/90/01}  % Allows default copyright data (0-89791-88-6/97/05) to be over-ridden - IF NEED BE.
% --- End of Author Metadata ---

\title{\underline{RE}\underline{SP}\underline{ARC}: A \underline{Re}configurable and Energy-Efficient \underline{Arc}hitecture with Memristive Crossbars for Deep \underline{Sp}iking Neural Networks}

\author{
%% You can go ahead and credit any number of authors here,
%% e.g. one 'row of three' or two rows (consisting of one row of three
%% and a second row of one, two or three).
%%
%% The command \alignauthor (no curly braces needed) should
%% precede each author name, affiliation/snail-mail address and
%% e-mail address. Additionally, tag each line of
%% affiliation/address with \affaddr, and tag the
%% e-mail address with \email.
%%
% 1st. author
\alignauthor
{Aayush Ankit, Abhronil Sengupta, Priyadarshini Panda, Kaushik Roy}%\titlenote{Dr.~Trovato insisted his name be first.}
      \\
       \affaddr{School of Electrical and Computer Engineering, Purdue University}\\
      %\affaddr{1932 Wallamaloo Lane}\\
       %\affaddr{}\\
       \email{\{aankit, asengup, pandap, kaushik\}@purdue.edu}
}

\maketitle
\begin{abstract}
Neuromorphic computing using post-CMOS technologies is gaining immense popularity due to its promising abilities to address the memory and power bottlenecks in von-Neumann computing systems. In this paper, we propose RESPARC - a reconfigurable and energy efficient architecture built-on Memristive Crossbar Arrays (MCA) for deep Spiking Neural Networks (SNNs). Prior works were primarily focused on device and circuit implementations of SNNs on crossbars. RESPARC advances this by proposing a complete system for SNN acceleration and its subsequent analysis. RESPARC utilizes the energy-efficiency of MCAs for inner-product computation and realizes a hierarchical reconfigurable design to incorporate the data-flow patterns in an SNN in a scalable fashion. We evaluate the proposed architecture on different SNNs ranging in complexity from 2k--230k neurons and 1.2M--5.5M synapses. Simulation results on these networks show that compared to the baseline digital CMOS architecture, RESPARC achieves 500$\times$ (15$\times$) efficiency in energy benefits at 300$\times$ (60$\times$) higher throughput for multi-layer perceptrons (deep convolutional networks). Furthermore, RESPARC is a technology-aware architecture that maps a given SNN topology to the most optimized MCA size for the given crossbar technology.
%This paper provides a sample of a \LaTeX\ document which conforms,
%somewhat loosely, to the formatting guidelines for
%ACM SIG Proceedings. It is an {\em alternate} style which produces
%a {\em tighter-looking} paper and was designed in response to
%concerns expressed, by authors, over page-budgets.
%It complements the document \textit{Author's (Alternate) Guide to
%Preparing ACM SIG Proceedings Using \LaTeX$2_\epsilon$\ and Bib\TeX}.
%This source file has been written with the intention of being
%compiled under \LaTeX$2_\epsilon$\ and BibTeX.
%
%The developers have tried to include every imaginable sort
%of ``bells and whistles", such as a subtitle, footnotes on
%title, subtitle and authors, as well as in the text, and
%every optional component (e.g. Acknowledgments, Additional
%Authors, Appendices), not to mention examples of
%equations, theorems, tables and figures.
%
%To make best use of this sample document, run it through \LaTeX\
%and BibTeX, and compare this source code with the printed
%output produced by the dvi file. A compiled PDF version
%is available on the web page to help you with the
%`look and feel'.
\end{abstract}
\vspace{-2mm}




%
%  Use this command to print the description
%

% We no longer use \terms command

\vspace{-2mm}

\vspace{-1mm}
\section{Introduction and Related Work}
Deep Learning Networks (DLN) inspired from the hierarchical organization of neurons and synapses in human brain are an important class of machine learning algorithms and have redefined the state-of-the-art for many cognitive applications \cite{krizhevsky2012imagenet}. However, DLNs involve data-intensive computations that lead to high power and memory bandwidth requirements on von-Neumann machines. As a result, the power budget they thrive on is multiple orders of magnitude greater than the human brain. For instance, AlexNet \cite{krizhevsky2012imagenet} that won the ImageNet challenge in 2012 consisted of 650k neurons and 60M synapses, and thrives on 2-4 GOPS of compute power per classification. Such power and memory bottlenecks have inspired the research in neuromorphic computing to build efficient architectures for accelerating neural networks by overcoming the von-Neumann bottlenecks. To this effect, several works have shown DLN implementations using graphic processing units, multi-core processors and hardware accelerators \cite{chetlur2014cudnn, chen2014diannao}. 

While DLNs are being successfully used in many recognition applications, there is a growing shift in the research community towards a more biologically plausible and energy-efficient computing paradigm, Spiking Neural Networks (SNN) \cite{diehl2015fast}. Driven by brain-like spike based computations, SNN involves event-driven data processing making them the emerging choice for energy-efficient recognition applications. Additionally, recent researches have shown deep SNNs to exhibit high accuracy on various complex recognition tasks \cite{diehl2015fast}. However, CMOS implementations of neuromorphic systems to accelerate SNNs suffer from power and area inefficiencies that stem from the realization of neuron and synapse functionality using primitives namely instructions and Boolean logic, resulting in dozens of transistors to mimic a single neuron/synapse \cite{akopyan2015truenorth}.

The limitations of CMOS can be addressed with emerging technologies, such as memristive devices that realize synaptic functionalities with very high efficacy \cite{jo2010nanoscale}. Crossbars made up of these devices at the cross-points have been studied for energy-efficient inner-product engines \cite{prezioso2015training, liu2015reno}. This has furthered the efforts to realize in-memory processing based architectures using Memristive Crossbar Arrays (MCA) for neuromorphic applications. However, MCA size is a strong function of technology, for example, Phase Change Memories (PCM) \cite{jackson2013nanoscale}, Ag-Si \cite{jo2010nanoscale}, Spintronic devices \cite{sengupta2016proposal} etc.  Large crossbars allow more flexibility in directly mapping an SNN onto it. This can also reduce peripheral overheads and thereby improve overall energy consumption. However, large crossbars are infeasible as they suffer from non-idealities like sneak-paths, process variations and parasitic voltage drops \cite{liang2010cross, liu2015vortex} which lead to erroneous computations. This necessitates the design of reconfigurable platforms for SNNs that can utilize the MCA energy benefits as well as address the limitations posed by MCA size.

In this work, we introduce RESPARC - a novel reconfigurable neuromorphic architecture built on MCAs for efficient implementation of SNN applications. The post-CMOS technology based MCAs provide efficient realization of synapses \cite{jo2010nanoscale}. Additionally, crossbars store the network weights thereby enabling ``in-memory processing''. This circumvents the problems associated with frequent and large volumes of data transfer between CPU and memory for implementing DLNs on conventional computing systems \cite{chi2016prime}. We translate the event-driven nature of SNN computations to architectural techniques (discussed in section 3.2) in order to achieve higher energy-efficiency. Hence, RESPARC aims at synergically combining the benefits of SNN and the design space of emerging technology using MCAs. Organizationally, RESPARC is a three-tiered reconfigurable platform designed to incorporate the data-flow patterns in any neural network in a scalable fashion. Each tier is targeted to bring in a specific variety of reconfigurability with respect to the SNN morphology. The three tiers are namely:
\vspace{-2mm}
\begin{enumerate} 
\item \textbf{Macro Processing Engine - reconfigurable compute unit} to map neurons with variable fan-in.
\vspace{-2mm}
\item \textbf{NeuroCell - reconfigurable datapath} to map SNNs with varying inter and intra layer connectivities namely Multi-Layer Perceptrons (MLPs) and Convolutional Neural Networks (CNNs).
\vspace{-2mm}
\item \textbf{RESPARC - reconfigurable core} to map SNNs with varying size (number of layers).
\end{enumerate}
\vspace{-2mm}
RESPARC is a spatially scalable architecture. It enumerates the synapses across MCAs on different mPEs, with mPEs spread across different NeuroCells thereby, using more MCAs for mapping a larger spiking neural network. Additionally, RESPARC's reconfigurability enables the usage of variable MCA sizes for mapping a given SNN topology. Hence, for any given MCA technology, a size which is permissible by the technology constraints for proper operation can be chosen thereby, enabling ``technology-aware'' mapping of SNNs on RESPARC.

Prior work on MCAs for SNN implementations have primarily focused on device and circuit optimizations and do not involve architecture-level analysis \cite{sengupta2016proposal}. The benefits at device-level need to be preserved at system-level. Our work proposes a full-fledged MCA based reconfigurable neuromorphic architecture that can implement a wide variety of SNNs with varying complexity and topology, as required by an application. It also helps to perform system-level analysis of MCAs, as MCAs are not a drop-in replacement for existing computation cores in the CMOS SNN implementations.

Post-CMOS based architectures were also explored in \cite{liu2015reno, chi2016prime, shafiee2016isaac}. While these works propose architectures for artificial neural networks, RESPARC targets SNN and utilizes its event-drivenness for added energy benefits. Additionally, RESPARC is distinct micro-architecturally as it explores a spatially scalable design based on reconfigurable hierarchies. Moreover, our design obviates the use of energy hungry analog-digital conversions unlike  \cite{chi2016prime, shafiee2016isaac} thereby leading to energy reductions.

There has also been prior work on SNN acceleration using CMOS technologies. For instance, Akopyan et al. \cite{akopyan2015truenorth} proposed TrueNorth which uses low power 28 nm CMOS technology and asynchronous circuit designs. While our work is complementary to the effects of \cite{akopyan2015truenorth}, we explore post-CMOS technology for SNN acceleration. Moreover, to analyze the benefits of RESPARC with respect to CMOS accelerators, we implement an optimized CMOS based baseline. Additionally, other techniques such as asynchronous computation will complement the SNN acceleration on RESPARC.

In summary the key contributions of this work are:
\vspace{-2mm}
\begin{enumerate} 
\item \textbf{An efficient memristive crossbar based architecture for spiking neural networks} is designed to harness the energy-efficiency from in-memory processing and event-driven computation.
\vspace{-2mm}
\item \textbf{Different spiking network topologies (MLP, CNN) from different recognition applications} namely digit recognition, house number recognition and object classification are mapped onto RESPARC and analyzed for performance and energy benefits with respect to their digital CMOS implementations.
\vspace{-2mm}
\item \textbf{Different MCA sizes for different SNN topologies based on the limitations posed by the memristive technology} are explored to determine the optimum crossbar size for mapping a given network. 
\end{enumerate}
\vspace{-2mm}

%The \textit{proceedings} are the records of a conference.
%ACM seeks to give these conference by-products a uniform,
%high-quality appearance.  To do this, ACM has some rigid
%requirements for the format of the proceedings documents: there
%is a specified format (balanced  double columns), a specified
%set of fonts (Arial or Helvetica and Times Roman) in
%certain specified sizes (for instance, 9 point for body copy),
%a specified live area (18 $\times$ 23.5 cm [7" $\times$ 9.25"]) centered on
%the page, specified size of margins (1.9 cm [0.75"]) top, (2.54 cm [1"]) bottom
%and (1.9 cm [.75"]) left and right; specified column width
%(8.45 cm [3.33"]) and gutter size (.83 cm [.33"]).
%
%The good news is, with only a handful of manual
%settings\footnote{Two of these, the {\texttt{\char'134 numberofauthors}}
%and {\texttt{\char'134 alignauthor}} commands, you have
%already used; another, {\texttt{\char'134 balancecolumns}}, will
%be used in your very last run of \LaTeX\ to ensure
%balanced column heights on the last page.}, the \LaTeX\ document
%class file handles all of this for you.
%
%The remainder of this document is concerned with showing, in
%the context of an ``actual'' document, the \LaTeX\ commands
%specifically available for denoting the structure of a
%proceedings paper, rather than with giving rigorous descriptions
%or explanations of such commands.

\section{Background}
\vspace{-2mm}
\subsection{Spiking Neural Network}
\vspace{-2mm}
\begin{figure}[h]
\centering
\includegraphics[width = 3in]{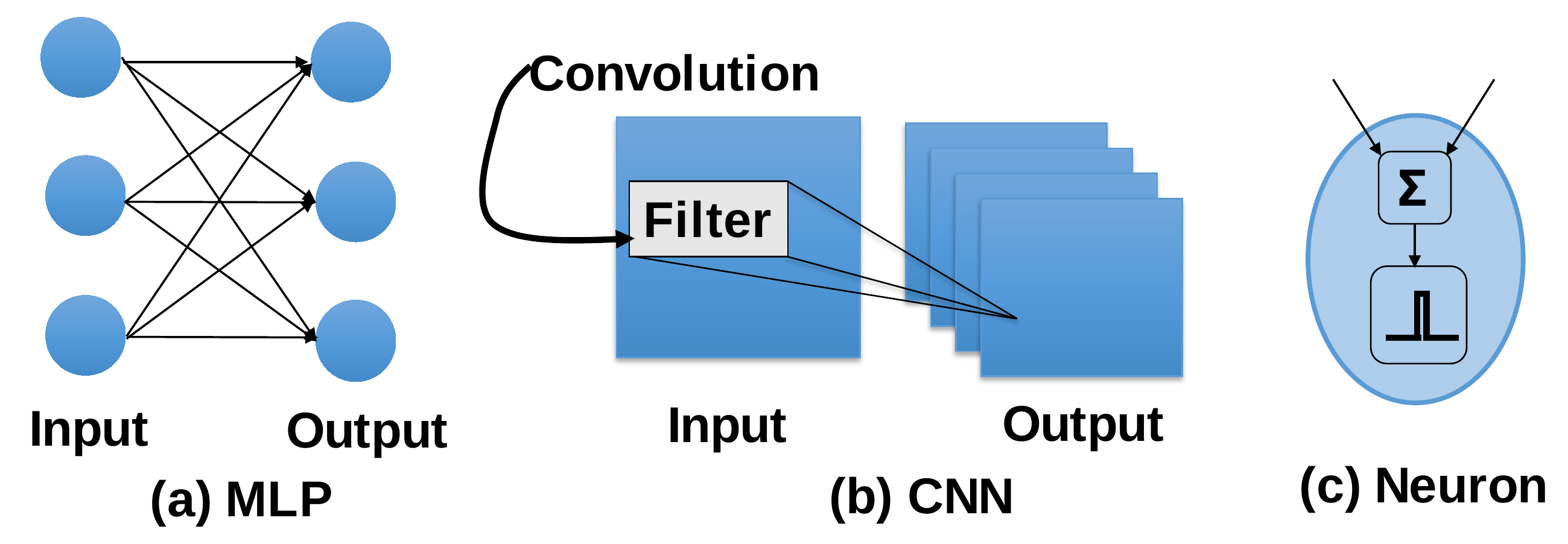}
\caption{\textbf{(a) A 2-layer MLP; (b) A convolution layer in CNN; (c) A Neuron}}
\label{fig_sim}
\end{figure}
\vspace{-2mm}

SNN is regarded as the third generation neural network. SNNs require the input to be encoded as spike trains and involve spike-based (0/1) information transfer between neurons. At a particular instant, each spike is propagated through the layers of the network while the neurons accumulate the spikes over time causing the neuron to fire or spike. The deep SNN topologies used in this work are MLPs and CNNs. An MLP, shown in Fig. 1(a), is a multi-layered SNN in which all neurons in a layer are connected to all neurons in the previous layer. A deep CNN, shown in Fig. 1(b), is also a multi-layered SNN composed of alternating convolution and sub-sampling layers. As shown in Fig. 1(c), a typical spiking neuron does an accumulation operation followed by thresholding operation. The spiking neuron model used in this work is the Integrate-and-Fire (IF) model. Note that, our work focuses on the testing/computation of the SNN and assumes that RESPARC has been trained offline using supervised training algorithms \cite{diehl2015fast}.

\subsection{Memristive Crossbars}
\vspace{-2mm}
\begin{figure}[h]
\centering
\includegraphics[width = 3in]{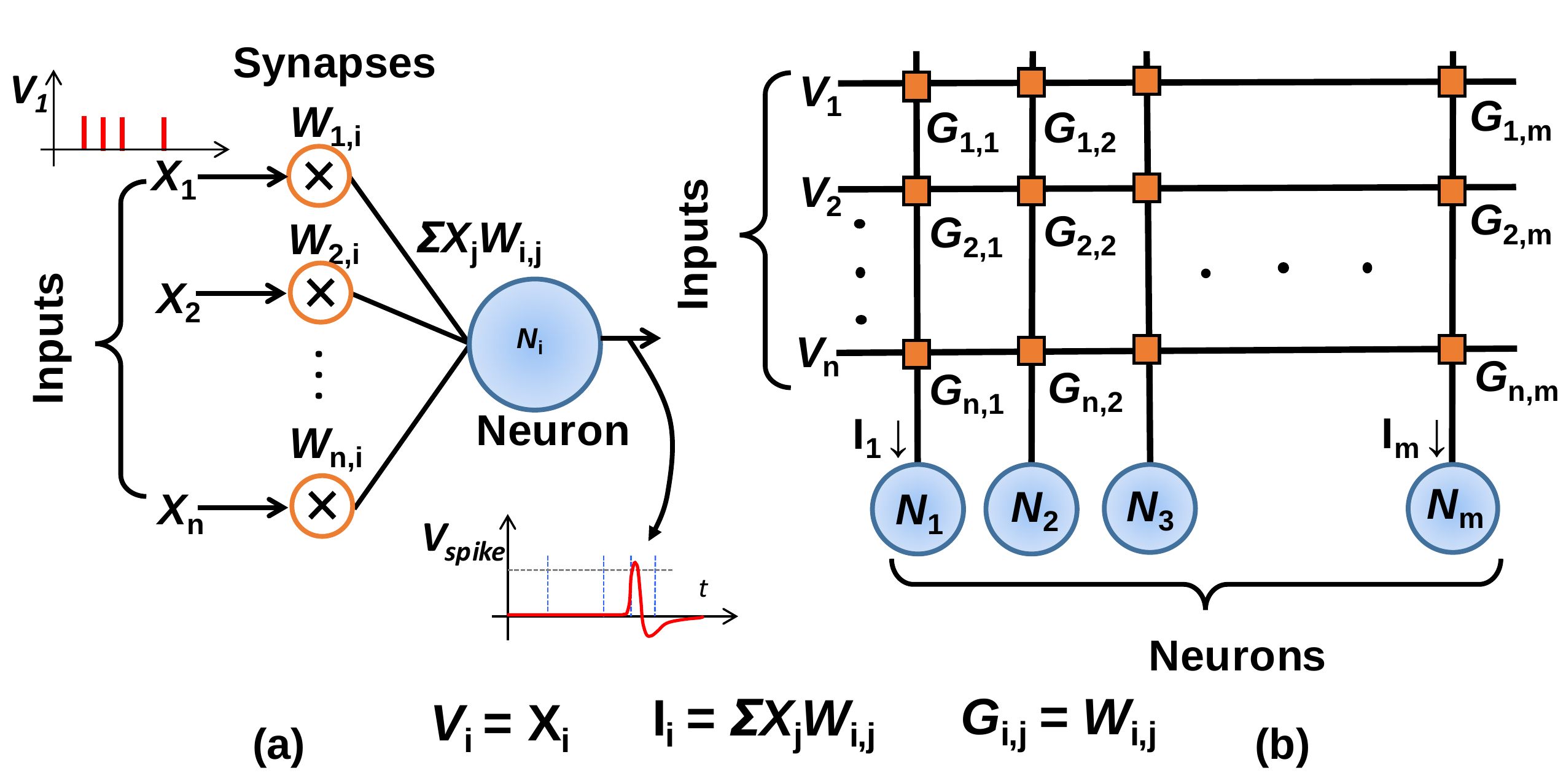}
\caption{\textbf{(a) Typical Spiking Neural Network (SNN) (b) SNN mapped to Memristive Crossbar Array (MCA)}}
\label{fig_sim}
\end{figure}
\vspace{-2mm}

Fig. 2(a) shows a 2-layer fully connected SNN. Fig. 2(b) shows the connectivity structure/matrix (from Fig. 2(a)) map-ped onto an MCA. The memristive devices at its cross-points encode the synaptic weights of the SNN. An MCA receives voltage inputs at its rows and the resulting current output at any column is the weighted summation of the encoded weights at that column and the input voltage. This is a direct consequence of the Kirchhoff's law as the current output into a column from any cross-point will be the product of the conductance at that cross-point and the voltage across it. Thus, MCA is an analog ``inner-product'' computation unit. The MCA outputs are interfaced with neurons. The neurons receive the input current that results in its membrane potential accumulating over time. When the membrane potential reaches a threshold, the neuron spikes (``1'') thus mirroring the function of an IF neuron.

%\vspace{-1mm}
\section {RESPARC}
\vspace{-2mm}
\subsection{Reconfigurable Hierarchies}
%\vspace{-1mm}
RESPARC (shown in Fig. 3) is the reconfigurable core and the topmost level among the three reconfigurable hierarchies. As shown in Fig. 3, RESPARC is composed of pool of NeuroCells which are the second level in the reconfigurable hierarchy. Fig. 4 shows a macro Processing Engine (denoted as mPE in Fig. 3) which is the lowest level in the hierarchy. Next, we will discuss the organization and the logical dataflow in each hierarchy starting from the lowest level and moving towards the higher levels.
\begin{figure}[!t]
\centering
\includegraphics[width = 3in]{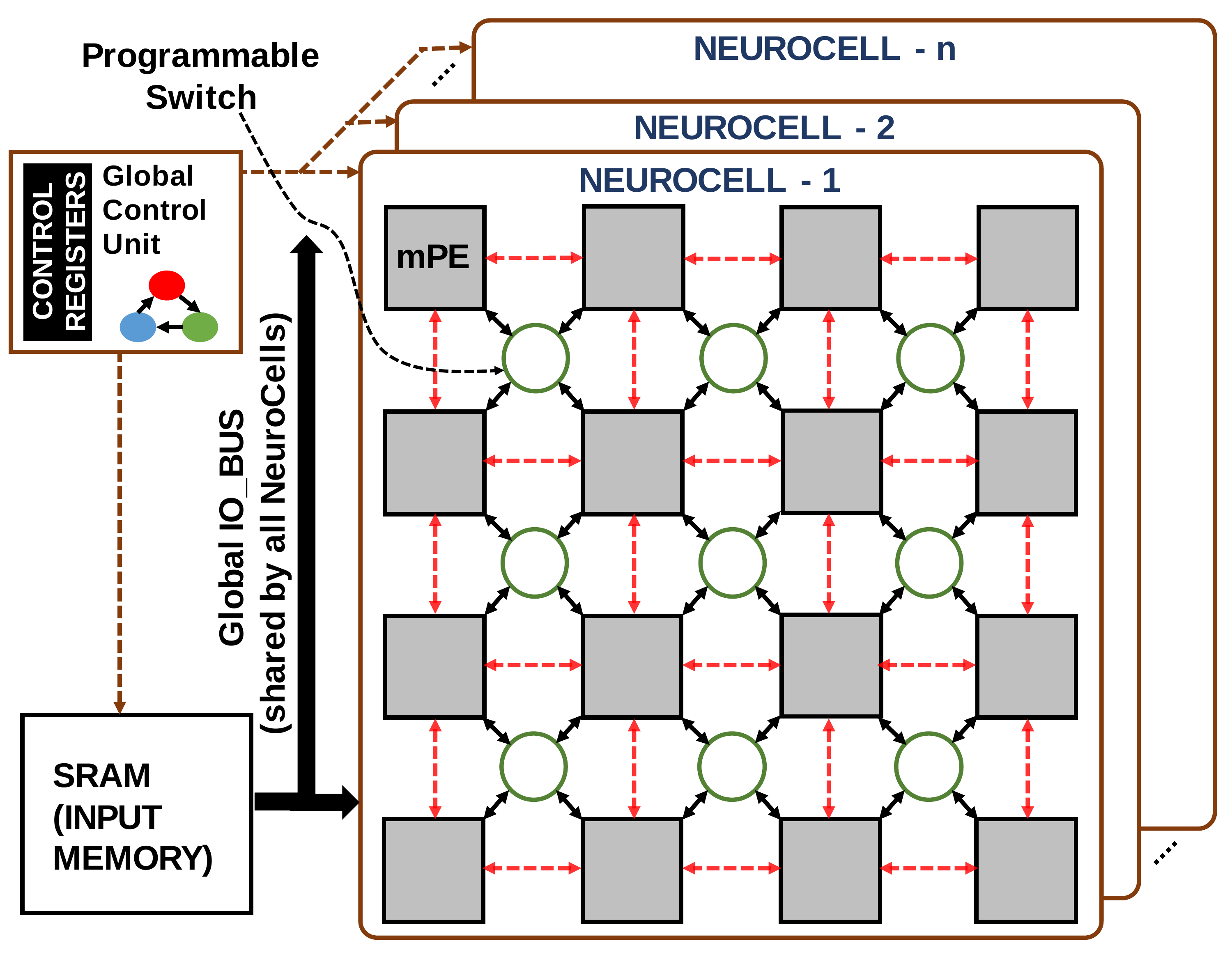}
\caption{\textbf{RESPARC as a pool of NeuroCells}}
\label{fig_sim}
\end{figure}
\setlength{\textfloatsep}{5pt minus 1.0pt}

\subsubsection{mPE - Reconfigurable Compute Unit}
A macro Processing Engine (mPE) is composed of multiple MCAs tied together to the Local Control Unit. The mPE shown in Fig. 4 includes four MCAs, each of which is associated to its neurons and a set of buffers namely (1) Input Buffer (iBUFF), (2) Output Buffer (oBUFF) and (3) Target Buffer (tBUFF). The iBUFF buffers the input spike packets received until the required data needed by the MCA is available. Similarly, the oBUFF buffers the output spike packets computed by the neuron until the required data to be sent to a target neuron is available. The tBUFF stores the address of the target neuron(s). Although, we consider IF neurons in this work, any spiking neuron can be interfaced with the MCA.

The MCAs contain the synapses corresponding to the neurons being computed in an mPE. This is realized by mapping the connectivity matrix on the MCAs as shown in Fig. 2. However, for memristive technology, MCA sizes which ensure reliable operation are much smaller for instance 64 rows and 64 columns (64 $\times$64) in comparison to a typical neural network's fan-in that is of the order of several hundreds \cite{liang2010cross}. This necessitates partitioning the connectivity matrix to map it across multiples MCAs. Subsequently, the neuron output is computed by time-multiplexing the MCA outputs onto the neuron as shown in Fig. 5. An mPE can be configured to support time-multiplexed computation of multiple degrees to map neurons with variable fan-in. In case a neuron's fan-in exceeds the fan-in support an mPE provides locally, the connectivity matrix is mapped across multiple mPEs.

\begin{figure}[!t]
\centering
\includegraphics[width = 3.4in]{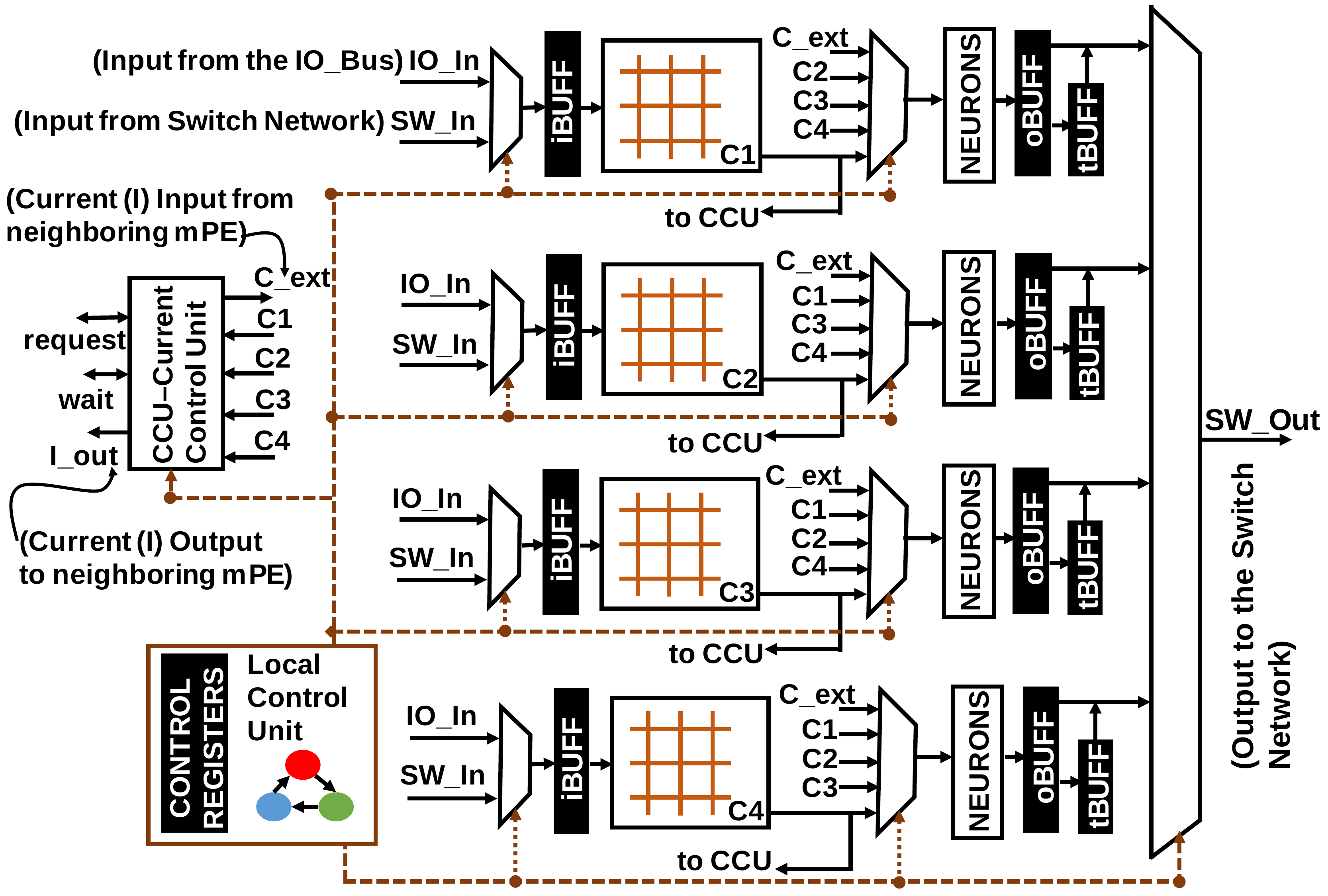}
\caption{\textbf{Macro Processing Engine - The mPE receives input spikes from the IO\_Bus and the Switch Network which is processed by the MCAs to produce output currents - C1, C2, C3, C4. Additionally, external MCA currents (C\_ext) can also be received by an MPE. Finally, the MCA currents get integrated into the neurons to produce output spikes that are then sent to the target neurons through Switch Network. The CCU controls the transfer of MCA currents to and fro between two mPEs.}}
\label{fig_sim}
\end{figure}

For sparser connectivity matrices, which is typical of CNNs, different output neurons have different inputs along with some input sharing. Hence, a column (column maps to an output neuron) in an MCA will consist of synapses at only certain sparse locations (rows) that correspond to its inputs leading to incompletely utilized MCA. Further, mapping the connectivity matrix of a CNN directly to a large MCA results in higher non-utilization due to large number of unused cross-points (synapses). However, enumerating the connectivity matrix across multiple smaller MCAs facilitates enhanced input-sharing that improves MCA utilization. Consequently, this reduces the number of mPEs required for the mapping. This improves overall energy consumption by reducing the peripheral energy per MCA. Hence, mPE's reconfigurability enables optimized MCA utilization for sparse connectivity.

\begin{figure}[!b]
\centering
\includegraphics[width = 3in]{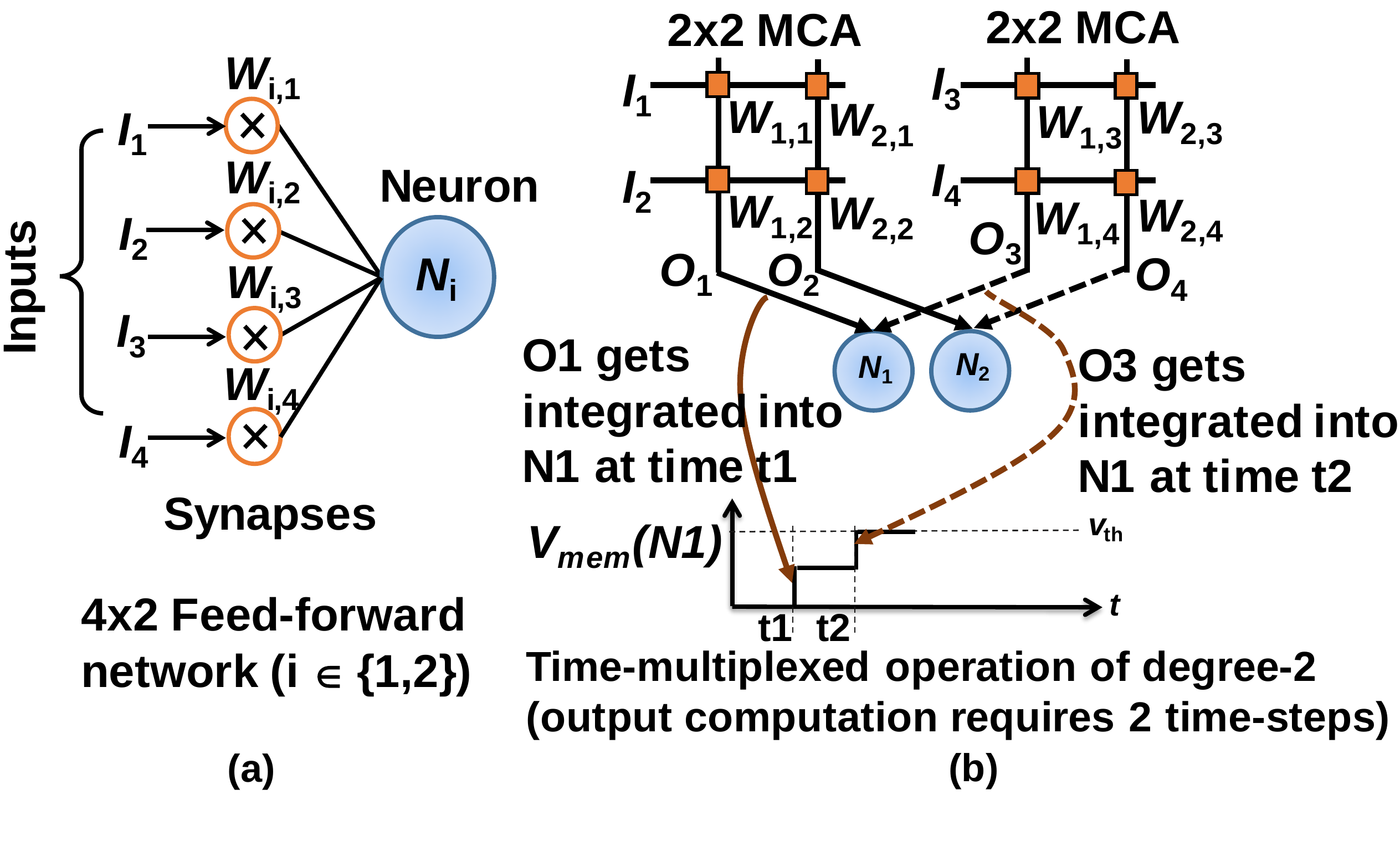}
\caption{\textbf{(a) A feed-forward neural network with neuron fan-in of 4 (b) Mapping the 4 fan-in neurons using a 2$\times$2 MCAs}}
\label{fig_sim}
\end{figure}

\subsubsection{NeuroCell - Reconfigurable Datapath}
As shown in Fig. 3, a NeuroCell is composed of multiple mPEs and programmable switches. The switch network enables spike-packet transfers within the NeuroCell. A switch connects to its four neighboring mPEs. Additionally, each switch has a dedicated connection to the switches in the same row and same column. This enables low-latency (one-hop) spike-packet transfers between the connected mPEs. Essentially, a NeuroCell is a pool of mPEs coupled with dense local connections that enables high throughput digital data transfer within it.  

Each switch can be configured to serve one or multiple mPEs it connects to thereby, realizing a reconfigurable datapath within the NeuroCell. This enables to optimize the datapath for the given SNN's connectivity. Consequently, this reduces the load on each switch and simplifies the overall traffic management within the NeuroCell. Fig. 6 shows the programmable switch design. Each input and output line is associated with data and address buffers to synchronize data transfer between the receiver and target mPE. Further, depending on the switch configuration, it arbitrates between the sender mPEs. 

As mentioned before, a connectivity matrix can span across multiple mPEs. To compute the neuron output, MCA current(s) from one mPE is transmitted to another mPE (consisting the neuron) followed by their time-multiplexed integration. Such analog signal transmission is facilitated by gated wires connecting the neighboring mPEs (dashed lines in Fig. 3).

\begin{figure}[!t]
\centering
\includegraphics[width = 3in]{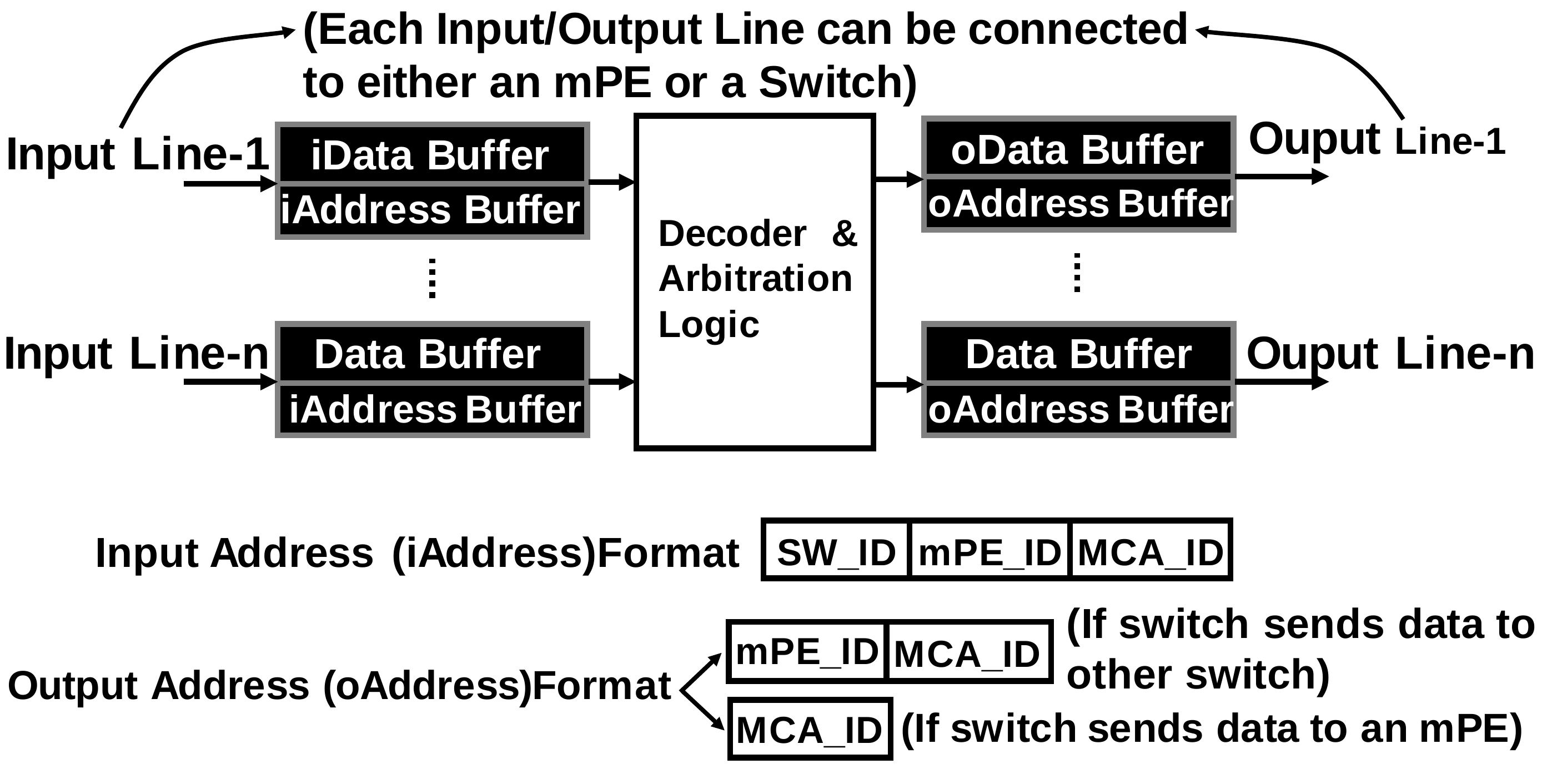}
\caption{\textbf{Programmable Switch}}
\label{fig_sim}
\end{figure}

\subsubsection{RESPARC - Reconfigurable Core}
RESPARC is the scalable extension of a NeuroCell (NC) and enables mapping of an SNN (that exceeds an NC's size) across multiple NCs. The NCs share a global ``IO\_BUS'' that connects to an SRAM (Input Memory). Thus, data transfer between different NCs go through the SRAM. Each NC in the NC-array is associated with a ``tag (x, y)'' which facilitates input broadcast from the SRAM to a variable number of NCs (that map to a given layer) within a single cycle. To monitor the completion of an NC's computation, the global control unit consists of an event-flag, dedicated to every NC which gets set when the NC completes.

\vspace{-2mm}
\begin{figure}[h]
\centering
\includegraphics[width = 3in]{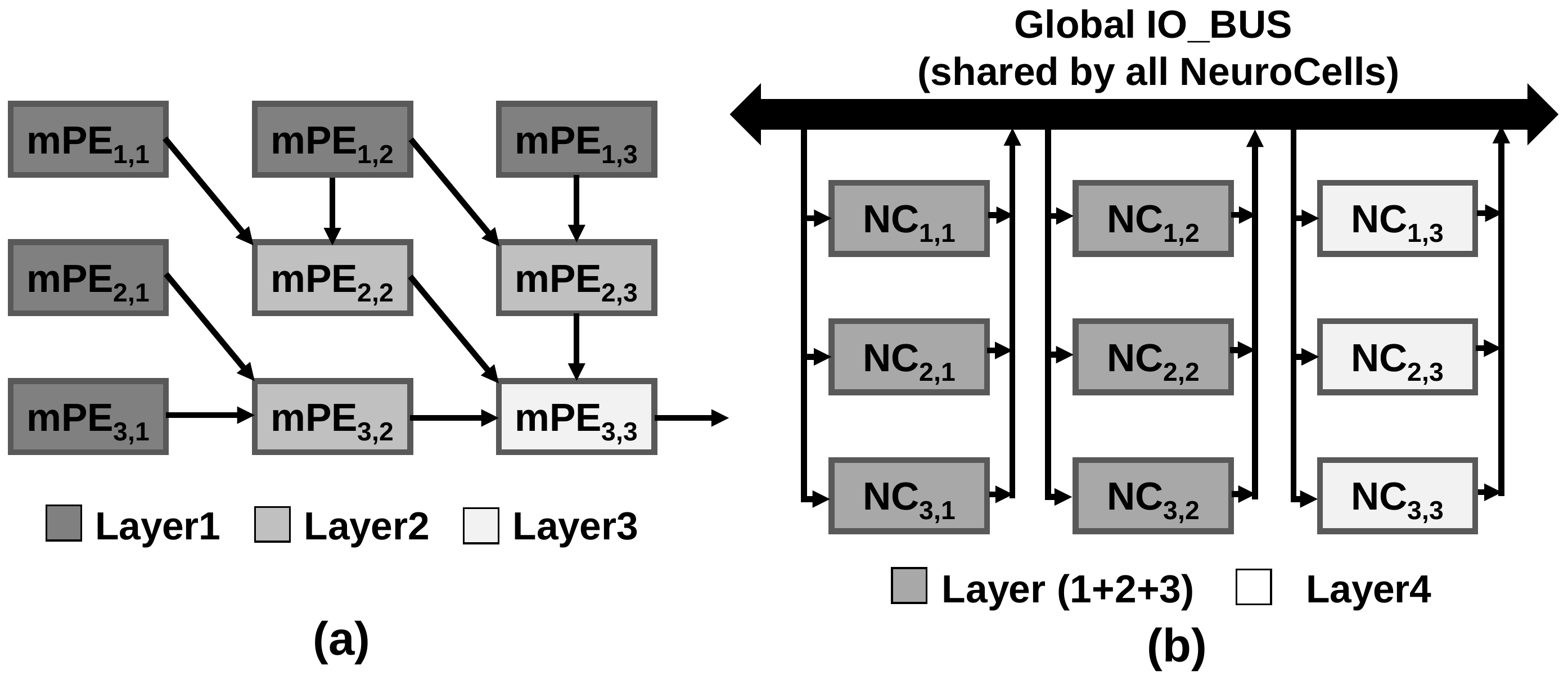}
\caption{\textbf{(a) Logical Dataflow in NeuroCell - High Throughput Data Transfer across switch network (multiple mPEs send data to multiple mPEs in parallel) (b) Logical Dataflow in RESPARC - Serial Data Transfer across shared bus}}
\label{fig_sim}
\end{figure}
\vspace{-2mm}

Fig. 7 illustrates the logical dataflow involved across hierarchies for SNN computation. Within an NC, parallel data transfer occurs between layers of the SNN through the switch network. Data transfer occurs serially through the shared bus between layers mapped across multiple NCs to compute the final output.

\subsection{Energy Efficiency}
We leverage the energy-efficiency of MCA (weight storage and inner-product computation) for energy savings. Additionally, as mentioned in section 3.1, the reconfigurability in mPE enables optimized mapping that reduces the peripheral energy per MCA thereby resulting in overall energy reductions. Within a NeuroCell, event-drivenness in SNN computations is utilized by adding ``zero-check logic'' in each programmable switch to prevent data transfers resulting from insignificant spike-packets (for instance, all bits in the spike packet being zero). Additionally, at the topmost level in the hierarchy, RESPARC exploits SNN data statistics (event-drivenness) to prevent unnecessary broadcasts to NeuroCells by checking the data read from SRAM with a ``zero-check logic''. Thus, the reconfigurability and event-driven computation further complement the benefits observed with MCAs for energy-efficient SNN acceleration on RESPARC.

\vspace{-1mm}
\section{Experimental Methodology}
\vspace{-1mm}
\subsection{CMOS Baseline}
We implemented the dataflow proposed in \cite{panda2016falcon} for our CMOS baseline and aggressively optimized it for SNNs. We augmented the implementation with event-driven optimizations to prevent unnecessary memory fetches and computations. Additionally, we added buffers to optimize the temporal and spatial data reuse patterns to minimize the memory fetches and thereby, optimizing the overall energy consumption. Note that our CMOS baseline enables to decouple the circuit and network-on-chip driven optimizations in other CMOS based SNN accelerators in order to rigorously analyze the MCA centric memory and computation benefits in RESPARC.

\subsection{Architecture Level Simulation Setup}
\begin{figure}[!b]
\centering
\includegraphics[width = 2.8in]{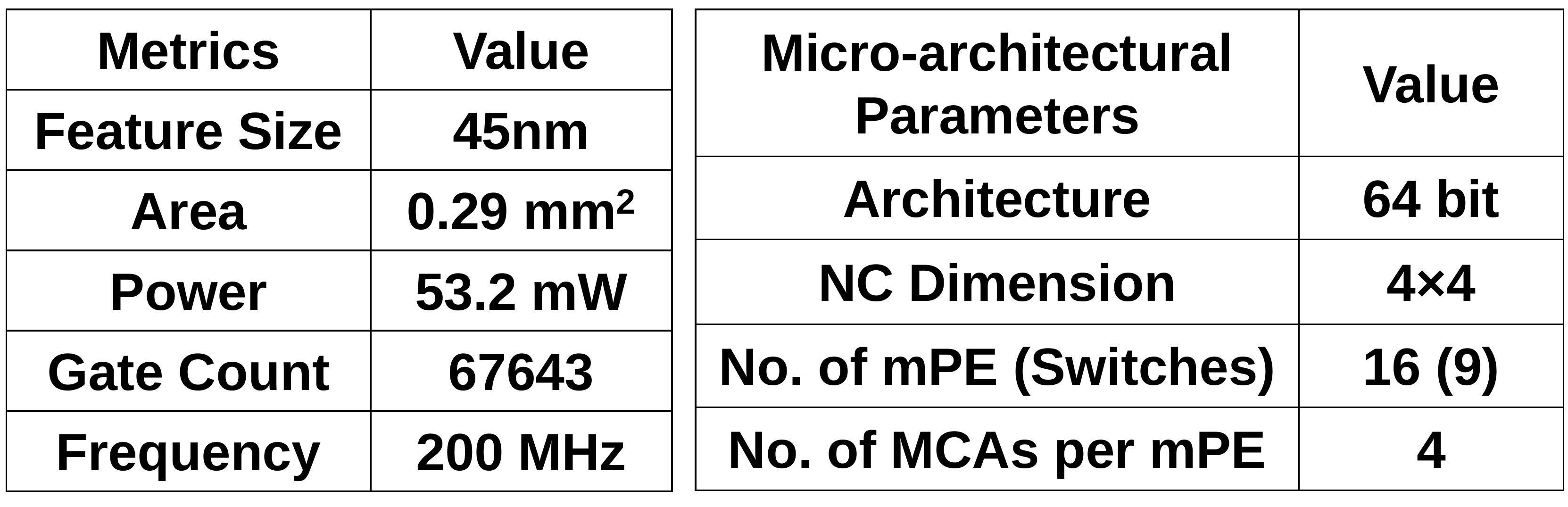}
\caption {\textbf{RESPARC parameters and metrics}}
\label{fig_sim}
\includegraphics[width = 2.8in]{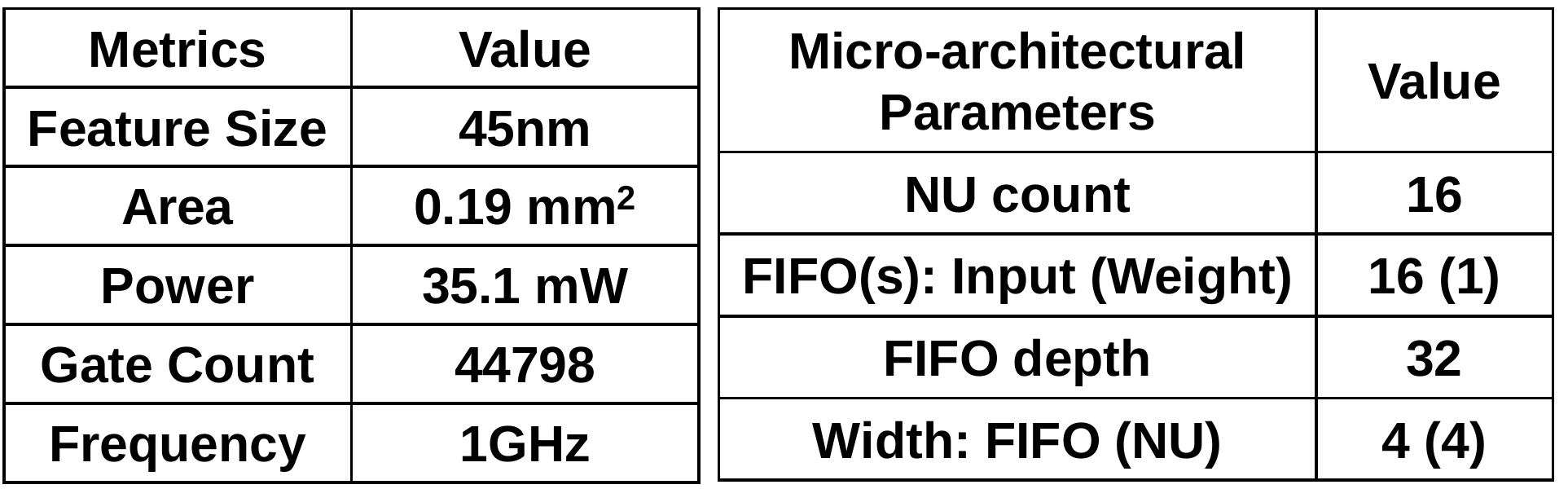}
\caption{\textbf{CMOS baseline parameters and metrics}}
\label{fig_sim}
\includegraphics[width = 2.8in]{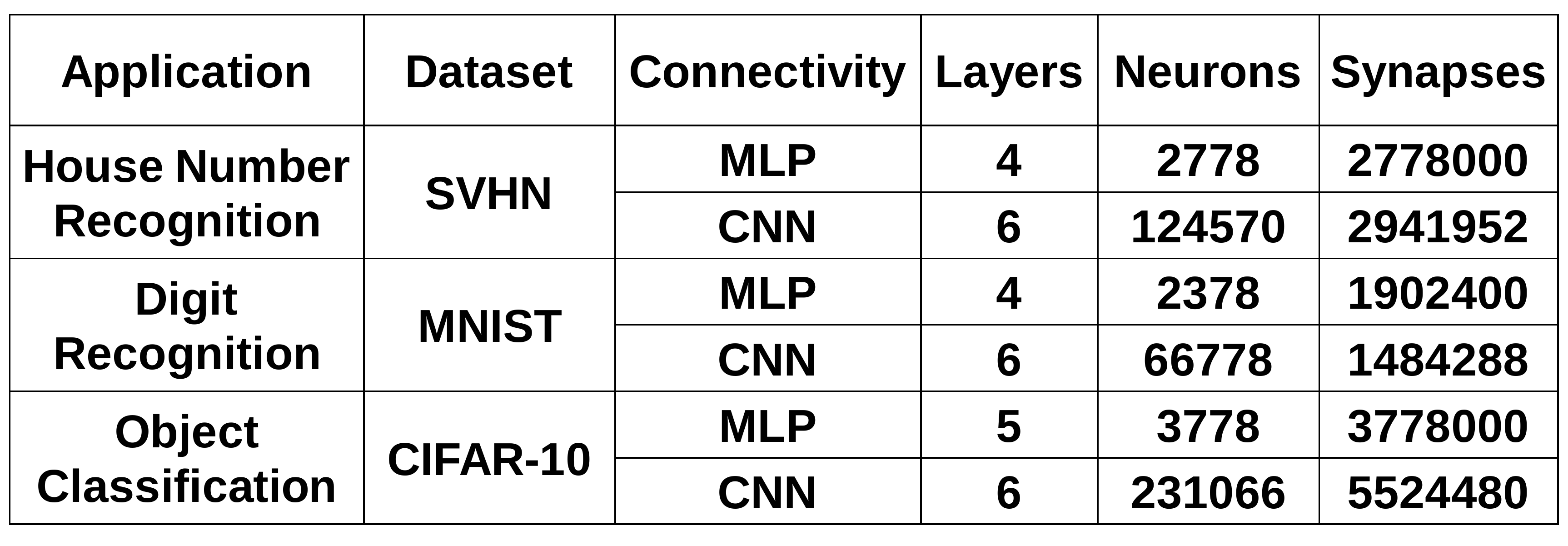}
\caption{\textbf{SNN Benchmarks}}
\label{fig_sim}
\end{figure}
%\vspace {-3mm}

RESPARC is composed of different technologies namely crossbar technology, technology of the interfaced neurons and the CMOS peripherals. For the memristive devices, we used a resistance range of ``20k$\Omega$ -- 200k$\Omega$'' with 16 levels (4 bits) for weight-discretization, that is typical of memristive technologies such as PCM, Ag-Si \cite{rajendran2013specifications}. We considered an operating voltage of ``Vdd/2'' for the MCA as it is interfaced with CMOS neurons \cite{joubert2012hardware}. The peripheral circuit consisting of buffers, communication and control logic was implemented at the Register Transfer Level in Verilog HDL and mapped to IBM 45nm technology using Synopsys Design Compiler. Synopsys Power Compiler was used to estimate the energy consumption. The input memory (SRAM) was modelled using CACTI  \cite{muralimanohar2007optimizing}. Fig. 8 lists the simulation parameters and the implementation metrics for one NeuroCell. Please note that the same methodology was also used to estimate the energy consumption of our CMOS baseline. Fig. 9 shows the simulation parameters and implementation metrics for the baseline. 

Our benchmark comprises of six SNN designs from different recognition applications namely, House Number Recognition (SVHN dataset \cite{netzer2011reading}), Digit Recognition (MNIST dataset\cite{lecun1998gradient}) and Object Classification (CIFAR-10 dataset \cite{krizhevsky2009learning}). We use one MLP and one CNN from each application. The SNNs were trained using supervised learning algorithm proposed in \cite{diehl2015fast}. Fig. 10 shows the benchmark details. As mentioned before, we do not consider the training phase of the SNN and hence, do not consider the energy expended in programming the MCAs. Also, in typical use case of recognition applications, the training process is performed once or very infrequently. On the other hand, the testing or evaluation phase, in which the actual classification is performed using SNNs, extends for much longer periods of time. Hence, we evaluate RESPARC for the more critical testing phase.

\vspace{-3mm}
\section {Experimental Results}
In this section, we present the results of various experiments that demonstrate the benefits of RESPARC and underscore the effectiveness of the proposed architecture in exploring the design space of post-CMOS based MCAs for SNN applications.

\vspace{-2mm}
\subsection {Comparison with CMOS baseline}
\begin{figure}[!b]
\centering
\includegraphics[width = 3.1in]{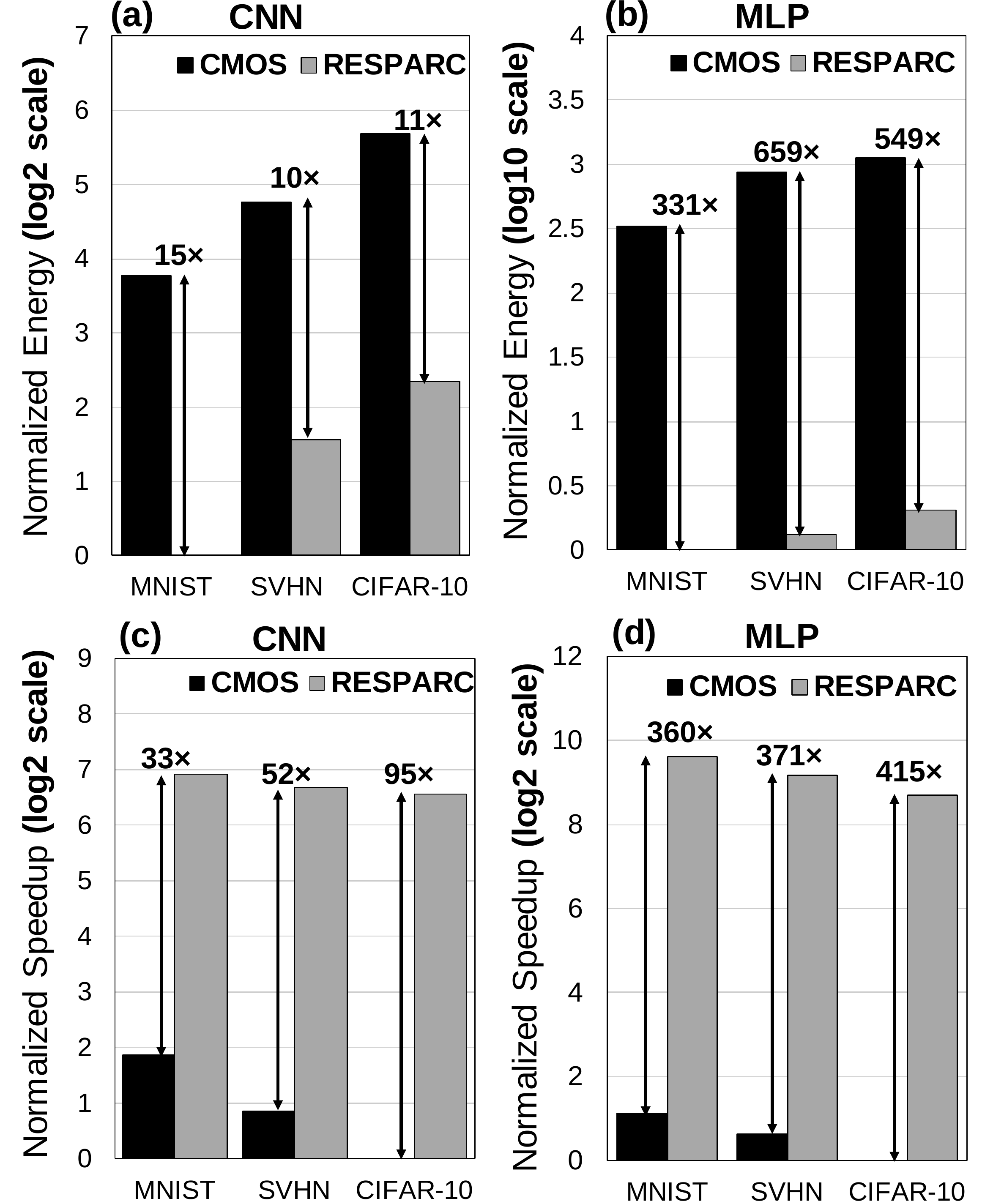}
\caption{\textbf{Energy and Performance Speedup comparison of RESPARC vs CMOS baseline per classification}}
\label{fig_sim}
\end{figure}
Fig. 11 compares the energy savings and performance spee-dups obtained \textit{per classification} for RESPARC over the CMOS baseline for various SNN applications with CNN and MLP topologies. The energy consumptions are normalized to the energy consumption of MNIST on RESPARC and the performance speedups are normalized to CIFAR-10 on CMOS baseline. The MCA size used is 64 i.e., 64 rows and 64 columns. As shown in Figs. 11 (a) and (c), RESPARC provides significant energy benefits between 10$\times$ -- 15$\times$ (12$\times$ on average) at a performance speedup of 33$\times$ -- 95$\times$ (60$\times$ on average) for the CNN benchmarks. For MLPs, (shown in Figs. 11(b) and (d)) energy benefits on RESPARC increase to 331$\times$ -- 549$\times$ (513$\times$ on average) at a performance speedup of 360$\times$ -- 415$\times$ (382$\times$ on average). Hence, RESPARC efficiently accelerates both CNN and MLP based SNN applications. 

The lower efficiency (both energy and speedup) for CNNs stems from the incomplete utilization of MCAs in RESPARC as discussed in subsection 3.1.1. The incomplete utilization leads to higher peripheral energy consumption per MCA there-by, decreasing the overall energy improvement. Additionally, incompletely utilized MCAs lead to lesser gain in performance speedup as lesser number of MCA outputs (columns) are utilized. In contrast, MLPs have fully utilized MCAs that result in higher throughput (number of outputs computed per unit time) as all the columns of the MCA are being used for output computation.

\vspace{-2mm}
\subsection {Comparision with varying MCA sizes}
The graphs in Fig. 12 show the breakdown of energy from Fig. 11 into 3 key components for RESPARC: (i) Neuron (ii) Crossbar (iii) Peripherals and 3 key components for the CMOS baseline: (i) Core (ii) Memory Access (iii) Memory Leakage. We present the energy distribution for MLP and CNN benchmarks on different MCA (crossbar) sizes namely (i) RESPARC-128 (ii) RESPARC-64 (iii) RESPARC-32. Fig. 12 (a) shows RESPARC energy consumption for MLPs. The energy consumption decreases with increasing MCA size. This is due to the fact that for larger MCAs the synapses would be mapped across less number of mPEs that decreases the peripheral energy per MCA reducing the overall energy consumption. On the other hand, for CNNs (shown in Fig. 12(c)), RESPARC-64 is the most energy-efficient. We observe a decrease in energy from RESPARC-32 to RESPARC-64 with CNNs due to decrease in peripheral energy. However, increasing MCA size from 64 to 128 increases the MCA non-utilization (due to sparser connectivity in CNNs as discussed in section 3.1.1) that dominates the overall energy consumption. Hence, unlike MLPs, an increase in MCA size from 64 to 128 does not result in a corresponding decrease in the peripheral energy per MCA as the number of mPEs being used does not decrease commensurately.

\begin{figure}[!t]
\centering
\includegraphics[width = 3.5in]{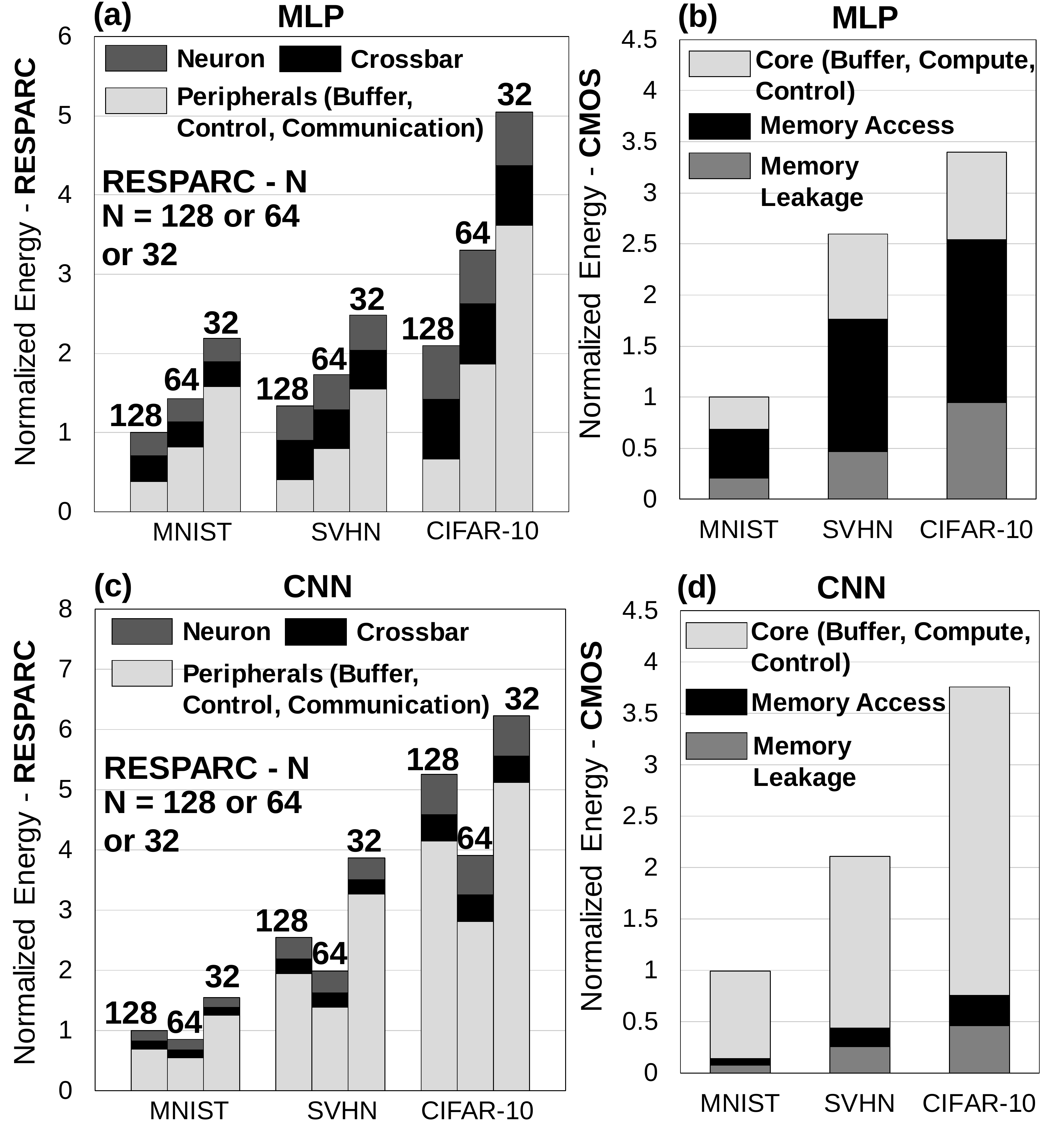}
\caption{\textbf{RESPARC and CMOS baseline energy breakdowns for different topologies}}
\label{fig_sim}
\end{figure}
\setlength{\textfloatsep}{4pt minus 1.0pt}

As shown in Fig. 12 (b), the energy consumption in MLPs on the CMOS baseline is dominated by the memory component (access and leakage). This implies that the energy savings for MLPs on RESPARC results from efficient memory storage (weight storage in MCAs). On the other hand, Fig. 12 (d) shows that the computation core (which includes the buffers and the computation units) dominates the energy consumption in CNNs. This suggests that the energy efficiency for CNNs on RESPARC results from the efficient inner-product computation in the MCAs.

\vspace{-1mm}
\subsection {Effect of event-drivenness in SNNs}
The graphs in Fig. 13 show the energy savings for MNIST dataset on RESPARC due to SNN's event-driven processing nature. The energy benefits are highest on RESPARC with the smallest MCA size. This is a consequence of the fact that the probability of finding zeros with smaller run-lengths (zeros with run length of 32 refers to a 32-bit spike-packet with all bits being zero) is significantly higher than that with larger run-lengths. We also obtained similar energy improvements with event-driven optimizations on the other two datasets. As discussed before, smaller MCAs are preferred because of reliability but they suffer from increased peripheral energy consumption. However, RESPARC with its event-drivenness enables using MCAs of smaller sizes for efficient acceleration of SNNs.

\vspace{-3mm}
\begin{figure}[h]
\centering
\includegraphics[width = 3.4in]{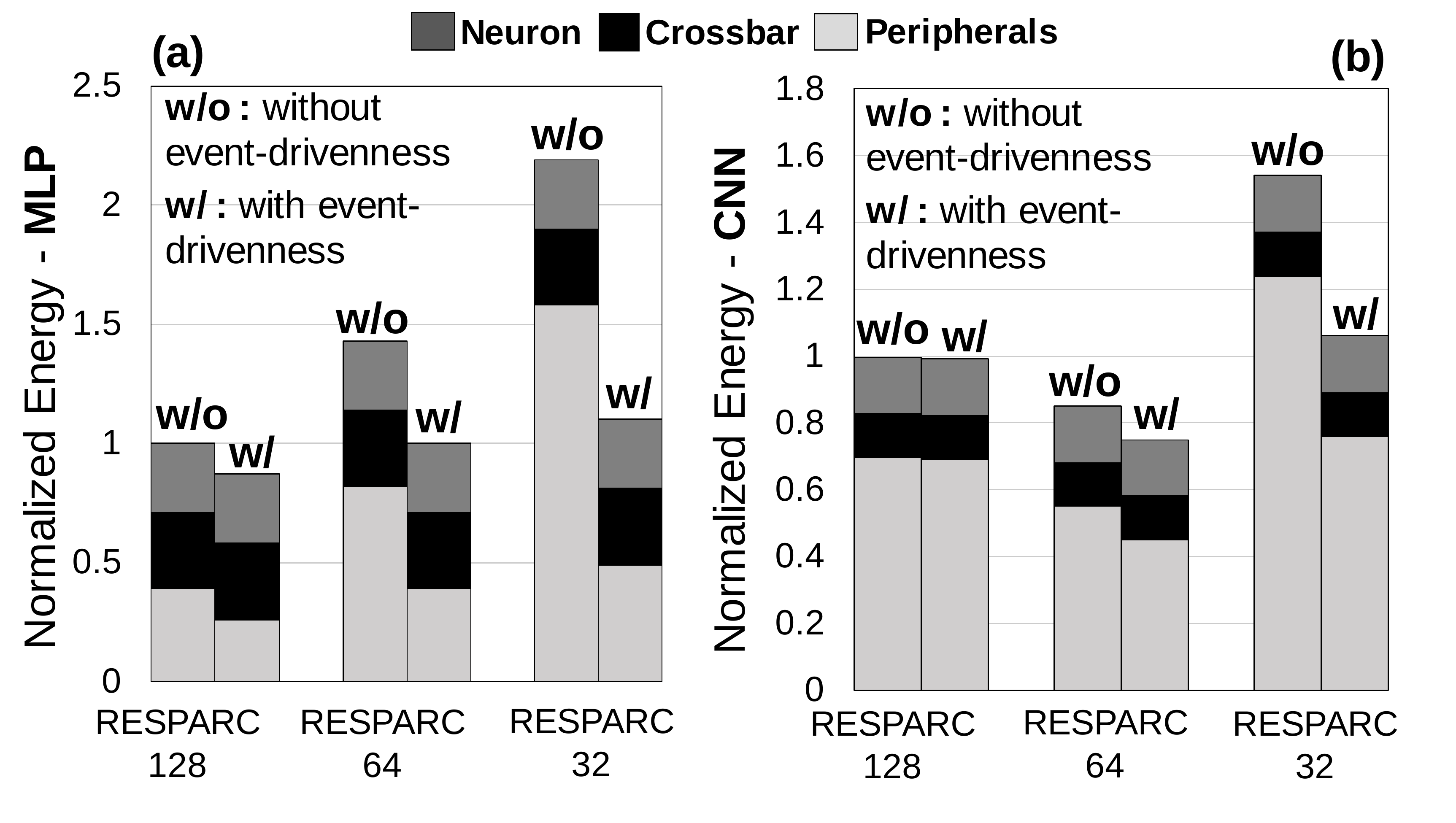}
\caption{\textbf{(a) Energy consumption analysis with event-drivenness in MLPs (b)  Energy consumption analysis with event-drivenness in CNNs}}
\label{fig_sim}
\end{figure}
\vspace{-3mm}

The benefits observed with CNNs are lesser than MLPs. This is due to the fact that CNNs process two-dimensional spatial windows of the input image that typically comprises of foreground (white) pixels. In contrast, MLPs process one-dimensional vectors that can easily find zero run-lengths for background (black) pixels.

\vspace{-1mm}
\subsection {Effect of bit-discretization of MCA}
Here, we analyze the memristor bit-precision on accuracy and energy consumption of RESPARC and CMOS baseline. As illustrated in Fig. 14 (a), the classification accuracy increases continuously with increasing weight precision (higher bit-discretization). However, the accuracy with 4-bits is comparable to the accuracy with 8 bits. Hence, we used 4-bit weight precision for our energy comparisons between RESPARC and CMOS baseline. However, other complex applications may necessitate the usage of higher bit-discretization for weight storage.

\vspace{-3mm}
\begin{figure}[h]
\centering
\includegraphics[width = 3.2in]{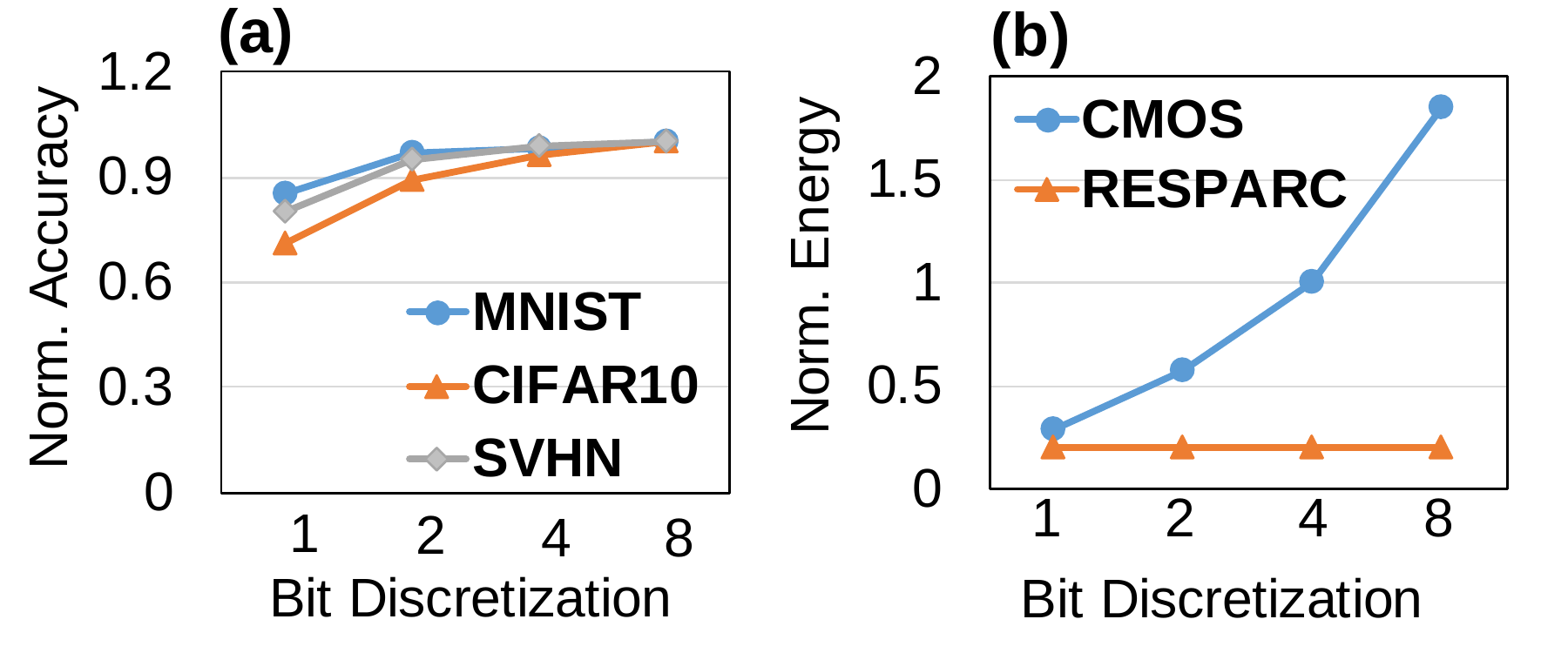}
\caption{\textbf{(a) Normalized Accuracy with respect to bit-dicretization in memristors (b) Normalized Energy with respect to bit-dicretization in memristors}}
\label{fig_sim}
\end{figure}
\vspace{-2mm}

A noteworthy observation here is that the energy consumption in RESPARC (from Fig. 14 (b)) is fairly independent of the weight precision. However, the area of the memristive device will increase with increasing precision that will increase the MCA area resulting in an area overhead. We also observe from Fig. 14 (b) that the energy consumption of the CMOS baseline increases with increasing bit-discretization. This is due to the fact that a higher precision demands bigger memory, buffers and compute units resulting in an increase in both the core power (buffer and computation units) and memory power (access and leakage).

\vspace{-2mm}
\section{Conclusions}
The intrinsic compatibility of post-CMOS technologies with biological primitives provides new opportunities to develop efficient neuromorphic systems. In this work we proposed RESPARC a memristive crossbar based architecture for energy-efficient acceleration of deep Spiking Neural Networks (SNN). We developed a reconfigurable hierarchy that efficiently implements SNNs of different connectivities given a memristive crossbar size and technology. Additionally, RESPARC synergically combines the energy benefits of post-CMOS technologies and the event-drivenness of bio-inspired SNNs to address the power and memory bottlenecks in modern computing systems. Our results on a range of recognition applications suggest that RESPARC is a promising architecture to implement SNNs providing favorable tradeoffs between energy and crossbar size.

\vspace{-2mm}
\bibliographystyle{abbrv}

\end{document}